\documentclass[%
 reprint,
showpacs,
nofootinbib,
 amsmath,amssymb,
 aps,
prd,
notitlepage,
twocolumn
]{revtex4-1}

\usepackage{graphicx}% Include figure files
\usepackage{dcolumn}% Align table columns on decimal point
\usepackage{bm}% bold math

\usepackage{amstext,amssymb,amsfonts,bbm,euscript,color,dsfont,mathtools}
\usepackage[utf8]{inputenc}
\usepackage{epsfig}
\usepackage{hyperref}
\usepackage{amsthm}
\usepackage{color}
\usepackage{xcolor}
\usepackage{multirow}
\usepackage{cancel}
\usepackage{dsfont}
\usepackage{mathrsfs}
\usepackage{physics}
\usepackage{verbatim} 
\usepackage{graphicx}
\usepackage{latexsym}
\usepackage{array}
\usepackage{cleveref}
\usepackage{subcaption}
\usepackage{ulem}
\usepackage{amsmath}

\newcommand{\bea}{\begin{eqnarray}}	
\newcommand{\eea}{\end{eqnarray}}
\newcommand{\be}{\begin{equation}}	
\newcommand{\ee}{\end{equation}}
\newcommand{\beq}{\begin{equation}}	
\newcommand{\eeq}{\end{equation}}

\newcommand{\IN}{{\mathbb N}}
\newcommand{\Z}{{\mathbb Z}}
\newcommand{\C}{{\mathbb C}}

\newcommand{\braopket}[3]{\left\langle{#1}|{#2}|{#3}\right\rangle}

\def\R{\relax\ifmmode {\mathbb R}  \else${\mathbb R}$\fi}
\def\C{\relax\ifmmode {\mathbb C}  \else${\mathbb C}$\fi}
\def\Z{\relax\ifmmode {\mathbb Z}  \else${\mathbb Z}$\fi}
\def\N{\relax\ifmmode {\mathbb N}  \else${\mathbb N}$\fi}
\def\I{\relax\ifmmode {\mathbb I}  \else${\mathbb I}$\fi}

\begin{document}

\title{Oscillators with imaginary coupling:
spectral functions in quantum mechanics and quantum field theory
}

\author{Bruno~W.~Mintz}\email{bruno.mintz@uerj.br} 

\affiliation{UERJ $-$ Universidade do Estado do Rio de Janeiro,\\
Departamento de F\'isica Te\'orica, Rua S\~ao Francisco Xavier 524,\\
20550-900, Maracan\~a, Rio de Janeiro, Brasil}

%%%
\author{Itai~Y.~Pinheiro}\email{iypinheiro@yahoo.com.br} 

\affiliation{UERJ $-$ Universidade do Estado do Rio de Janeiro,\\
Departamento de F\'isica Te\'orica, Rua S\~ao Francisco Xavier 524,\\
20550-900, Maracan\~a, Rio de Janeiro, Brasil}

%%%

\author{Rui~Aquino}\email{ruiaquino@ictp-saifr.org} 

\affiliation{ICTP - South American Institute for Fundamental Research - Instituto de F\'isica Te\'orica da UNESP, \\
Universidade Estadual Paulista, Rua Dr. Bento Teobaldo Ferraz 271, 01140-070 S\~ao Paulo, Brasil}

\date{\today}% It is always \today, today,

\begin{abstract}

The axioms of Quantum Mechanics require that the hamiltonian of any closed system is self-adjoint, so that energy levels are real and time evolution preserves probability. On the other hand, non-hermitian hamiltonians with ${\cal{PT}}$-symmetry can have both real spectra and unitary time evolution. In this paper, we study in detail a pair of quantum oscillators coupled by an imaginary bilinear term, both in quantum mechanics and in quantum field theory. We discuss explicitly how such hamiltonians lead to perfectly sound physical theories with real spectra and unitary time evolution, in spite of their non-hermiticity. We also analyze two-point correlation functions and their associated K\"allen-Lehmann representation. In particular, we discuss the intimate relation between positivity violation of the spectral functions and the non-observability of operators in a given correlation function. Finally, we conjecture that positivity violation of some spectral functions of the theory could be a generic sign of the existence of complex pairs of energy eigenvalues (i.e., a ${\cal{PT}}$-broken phase) somewhere in its parameter space.

%\begin{description}
%\end{description}
\end{abstract}

\maketitle

%\tableofcontents

%%%%%%%%%%%%%%%%%%%%%%%%%%%%%%%%%%%%%%%%%%%%%%%%%%%%%%%%%%%%%%
\section{\label{sec:Intro}Introduction}
%%%%%%%%%%%%%%%%%%%%%%%%%%%%%%%%%%%%%%%%%%%%%%%%%%%%%%%%%%%%%%

Non-hermitian systems are emerging as feasible platforms to describe new phenomena on different scales. In Non-Hermitian topological photonics, the topology ensures the robustness of propagation of electromagnetic waves, while the non-Hermiticity allows for more methods of wave manipulation \cite{El-Ganainy2019, Feng2017}. In open quantum systems, where the dynamics of density matrices are described by a Lindblad master equation, the effective Hamiltonian of the quantum system is non-hermitian, where the non-hermitian part describes the dissipation of energy of the original system with the bath \cite{fazio2024}. Some further applications rely on lattices systems, where dissipation can be generated in different ways for each site, or synthetically building setups with asymmetrical hopping. In ultracold atoms, for instance, dissipation is generated with lasers that ensure the loss of particles in a controlled way.

An important theoretical tool to describe dissipative quantum systems is the modeling of locally non-conservative systems by effective non-Hermitian Hamiltonians \cite{Rotter2015, Bagarello2016,Michishita2020}. These types of Hamiltonians have several counterintuitive properties. Perhaps, one of the most striking is the appearance of non-Hermitian degeneracies \cite{Berry1998} known as exceptional points (EPs) \cite{Heiss1990,Kato1995}. When a non-Hermitian Hamiltonian continuously depends on external parameters, it could happen that, for certain values of the parameters, two or more eigenvalues coalesce to an EP. However, this is not a usual degeneracy, as observed in Hermitian systems. In an exceptional point, not only the eingenvalues coincide, but also the eigenvectors become linearly dependent \cite{Rotter2007}, reducing in this way the dimension of the subspace associated with the degenerated eigenvalue. This singularity of the Hilbert space has remarkable topological consequences \cite{Lee2009,Xu2016,Doppler2016,Zhang2021}.

These complex level crossings are of tremendous importance for non-hermitian physics, as, for instance, on the topological properties of gapless non-hermitian phases \cite{Kawabata2019}. In the case of gapped phases, topological or not, exceptional points are exactly the phase transition points, i.e., the region in parameter space where the complex gap closes. One example where this behavior occurs is in $\mathcal{PT}$-symmetric (or, equivalently, pseudo-hermitian) systems. These are characterized by spectra with real eigenenergies or complex conjugate pairs, which we call the $\mathcal{PT}$-symmetric region and broken $\mathcal{PT}$-symmetry region, respectively \cite{Bender_1998,Bender_2007_70,Bender2024}. The point in which the real eigenenergies split into complex conjugates, i.e. the point in which the $\mathcal{PT}$ symmetry is broken, is an exceptional point. 

In recent years, there has also been a growing interest in ${\cal PT}$-symmetric Quantum Field Theories. For example, the ghosts of the Lee model \cite{Lee_Model_1954} and of the Pais-Uhlenbeck model \cite{PaisUhlenbeck_1950} (as a prototype for higher derivative QFTs) have been rendered harmless by acknowledging that these theories are ${\cal PT}$-symmetric \cite{Bender_2005,Bender:2007wu}. Various ${\cal PT}$-symmetric Quantum Field Theories have been explored, not only for scalar fields but also for higher spin fields  \cite{Bender:2018pbv,Bender:1999ek,Romatschke:2022jqg,Lawrence:2023woz,Ai:2022olh,Ai:2022csx,Alexandre:2017foi,Alexandre:2018uol,Raval:2018kqg,Alexandre:2018xyy,Beygi_2019,Alexandre_2020,Alexandre_2022,Felski:2022dsx,kuntz2024unitarityptsymmetryquantum}. 

In this paper, motivated by the well known positivity violation of spectral functions in Yang Mills theories, we will work with the so called pseudo-hermitian systems \cite{Mostafazadeh_2002_431}. We choose a simplified model of two harmonic oscillators with an imaginary coupling, in quantum mechanics and in quantum field theory. These models present interesting features, such as $\mathcal{PT}$-symmetry breaking in its spectrum. However, we will only focus on the $\mathcal{PT}$-symmetric phase, i.e., we will not study the complex part of the spectrum of the model. This choice is because the region with real eigenenergies still presents  positivity violation in its spectral function. 

We fully characterize the spectrum of both models, explicitly showing the real eigenenergies. Also, by ensuring that all observables must respect the same symmetry of the hamiltonian (i.e. be pseudo-hermitian) during all the time evolution, we showed that it is possible to ensure the positivity of the spectral function. We are able to do this by building a metric operator both for the quantum mechanical and the quantum field theory model. 

This paper is organized as follows. In Sec. \ref{sec:PT-symm-review}, we review some basic properties of pseudo-hermitian quantum mechanics, which will be used throughout the text. In Sec. \ref{sec:spectral-functions}, we discuss two-point correlation functions and the positivity (or not) of their associated spectral functions and how this can be closely related to the pseudo-hermiticity of the hamiltonian. Next, in Sec. \ref{sec:coupled-oscillators}, we study in detail a pair of harmonic oscillators coupled through an imaginary bilinear term, so that the resulting hamiltonian is not hermitian, but rather pseudo-hermitian. We generalize this discussion to scalar quantum fields in Sec. \ref{sec:qft-model}. Finally, in Sec. \ref{sec:discussion}, besides summarizing our findings and pointing out possible directions for future work, we conjecture that the relation between positivity violation of the spectral function and a nontrivial metric in Hilbert space is possibly not an accident of the model we studied, but likely a quite generic feature of Quantum Field Theories.

%\newpage

%%%%%%%%%%%%%%%%%%%%%%%%%%%%%%%%%%%%%%%%%%%%%%%%%%%%%%%%%%%%%%
\section{\label{sec:PT-symm-review}A few aspects of pseudo-hermitian quantum mechanics}
%%%%%%%%%%%%%%%%%%%%%%%%%%%%%%%%%%%%%%%%%%%%%%%%%%%%%%%%%%%%%%

In order to define the Hilbert space for the states of a given quantum theory, one has to define not only a set of state vectors, but also an inner product  \cite{vonNeumannQMbook}. Assuming some inner product (which we call the {\it reference inner product}, and denote by the standard Dirac notation, \(\langle \cdot\vert \cdot\rangle\)) we define the adjoint of a generic operator \(\hat {\cal O}\) in the reference inner product as \(\hat {\cal O}^\dagger\) according to
\begin{equation}
    \langle \psi \vert \hat{\cal O} \phi\rangle = \langle \hat {\cal O}^\dagger \psi\vert \phi \rangle.
\end{equation}
We emphasize that we use this notation ('dagger') only for the reference inner product. If the operator obeys \( \hat {\cal O} = \hat {\cal O}^\dag\), i. e., if it is self-adjoint with respect to the reference inner product, we call it \textit{hermitian}.

A point which is not always discussed in standard textbooks of Quantum Mechanics is that the Dirac inner product is not the only possibility for the construction of the Hilbert space of a quantum theory. Indeed, depending on the dynamics of a system (i.e., its hamiltonian and the dynamical variables used to express it), the choice of an adequate inner product is crucial for a sensible quantum theory. According to the Dirac-von Neumann postulates of Quantum Mechanics \cite{vonNeumannQMbook,DiracQMbook}, transition probabilities and expectation values of observables are derived from an adequate inner product. Evidently, a poor choice of inner product for the Hilbert space can lead to an ill-defined quantum theory.

Now, consider a quantum system endowed with a non-hermitian Hamiltonian $\hat H$ (meaning \(\hat H \not = \hat H^\dag\)). This hamiltonian is said to be $\eta-$pseudo-hermitian (or simply $\eta-$hermitian) if it respects the following similarity transformation, 
\begin{align}
    \hat H^\dagger = \hat \eta \hat H \hat \eta^{-1}, \label{eq:pseudo-hermitian}
\end{align}
where the operator $\hat \eta$ is called the
metric operator \cite{Mostafazadeh_2002_431}.
For hermitian and positive $\hat\eta$, the spectrum of $\hat H$ is real \cite{Mostafazadeh_2010_07}. Note that the metric operator is not unique, in the sense that infinitely many operators can possibly implement (\ref{eq:pseudo-hermitian}).

It has been known for quite some time that non-hermitian hamiltonians can have real spectrum \cite{Bender_1998}. Even if this is the case, there are at least two evident potential problems with a non-hermitian hamiltonian (if used to described a closed system) if one keeps using the reference Hilbert space. First, the eigenvectors of a non-hermitian operator with different eigenvalues are not mutually orthogonal. Therefore, an energy measurement would not properly project the system in a state with well-defined energy. Second, being the generator of time evolution, a non-hermitian hamiltonian leads to a non-unitary time evolution operator and, therefore, to non-conservation of total probability.

These problems can be solved, however, by noting that both of them rely on the definition of the Hilbert space inner product. A complete and discrete set of eigenvectors of the hamiltonian (with real eigenvalues) may be used to define a new inner product so that such eigenvectors then constitute an orthonormal basis. In this new Hilbert space (with the new inner product), the hamiltonian is self-adjoint \cite{Mostafazadeh_2010_07}. 

Let us now define an inner product which renders the hamiltonian self-adjoint as \cite{Bender_2002_89,Mostafazadeh_2002_431,Mostafazadeh_2010_07}
\begin{eqnarray}\label{eq:eta-inner-product}
    {\langle\psi \vert\phi\rangle}_{\eta} &:=& \langle \psi \rvert \hat \eta\, \phi \rangle.
      \label{eq: eta-product}
\end{eqnarray}
We call this the $\eta$-inner product. Note that, although definition (\ref{eq:eta-inner-product}) is built upon the reference inner product, the $\eta-$inner product exists on its own. Indeed, it could even be considered as the reference inner product from the start. The reason to redefine the inner product through Eq.~(\ref{eq: eta-product}) is to build a convenient connection between the Hilbert spaces. 

A different inner product implies a different conjugation rule for operators. With the connection between the reference and the $\eta-$inner product given by (\ref{eq: eta-product}), it is convenient to define the $\eta-$conjugate $\hat A^\#$ of any operator $\hat A$ so that
\begin{eqnarray}
    \braket{\psi}{\hat A\phi}_\eta = \braket{\hat A^\#\psi}{\phi}_\eta, \label{eq:eta-conjugation-def}
\end{eqnarray}
which is equivalent to defining \cite{Mostafazadeh_2002_431}
\begin{eqnarray}\label{eq:eta-conjugation-operator}
    \hat A^\# = \hat\eta^{-1} A^\dagger \hat\eta.
\end{eqnarray}

It is immediate to show that for any $\hat H$ that satisfies (\ref{eq:pseudo-hermitian}) it follows that
\begin{eqnarray}\label{eq:H-eta-hermitian}
    {\langle\psi \vert\hat H \phi\rangle}_{\eta} &=& {\langle\hat H\psi \vert \phi\rangle}_{\eta},
\end{eqnarray}
that is, $\hat H^\#=\hat H$. 
This justifies calling $\hat H$ a ``$\eta-$hermitian operator''. From a different (and equivalent) point of view, one could even say that $\hat H$ is indeed hermitian, as long as we start considering \(\langle \cdot\vert \cdot\rangle_\eta\) as the reference inner product.

It is relation (\ref{eq:H-eta-hermitian}) that supports the reality of the energy levels (but only if the inner-product is positive-definite \cite{Mostafazadeh_2010_07}). Although (pseudo-)hermiticity implies real eigenvalues, the converse is not necessarily true, because the very notion of an adjoint operator (i.e. $\hat {\cal O}^\dagger$ or $\hat {\cal O}^\#$) depends on the definition of a somewhat arbitrary inner product, whereas the spectrum of an operator can be defined regardless of any inner product.

Considering a generic (not necessarily positive) metric, the eigenvalue problem for the hamiltonian $\hat H$ is 
\begin{eqnarray}\label{eq:eigenvalue_eqn_H}
    \hat H\ket{\psi_n} = E_n\ket{\psi_n},
\end{eqnarray}
whereas the eigenvalue problem for $\hat H^\dagger$ is 
\begin{eqnarray}\label{eq:eigenvalue_eqn_H_dagger}
    \hat H^\dagger\ket{\phi_n} = E_n^*\ket{\phi_n}.
\end{eqnarray}
That is, $\ket{\psi_n}$ are the right-eigenvectors of $\hat H$, whereas $\ket{\phi_n}$ are its left-eigenvectors.
The $\eta-$hermiticity relation (\ref{eq:pseudo-hermitian}) implies that 
\begin{eqnarray}\label{eq:phi_eta_psi}
\ket{\phi_n}=\hat\eta\ket{\psi_n}.
\end{eqnarray}
Furthermore,
\begin{eqnarray}\label{eq:biorthonormality}
    \braket{\phi_n}{\psi_m}=\braket{\psi_n}{\hat\eta\psi_m}= \braket{\psi_n}{\psi_m}_\eta=\delta_{mn},
\end{eqnarray}
so that the set $\{\{\ket{\psi_n}\},\{\ket{\phi_n}\}\}$ is a biorthonormal basis for the Hilbert space. The formalism called Biorthogonal Quantum Mechanics \cite{Brody_2013} (which is equivalent to the pseudo-hermitian  formalism employed here) is constructed around this property of pseudo-hermitian hamiltonians.

The biorthonormality property (\ref{eq:biorthonormality}) leads to the representation of the identity operator as
\begin{eqnarray}\label{eq:completeness}
 \hat{\mathds{1}} = \sum_n\ket{\psi_n}\bra{\phi_n} = \sum_n\ket{\psi_n}\bra{\psi_n}\hat\eta,
\end{eqnarray}
since $\bra{\phi_n}=\bra{\psi_n}\hat\eta$.
Of course, for a hermitian hamiltonian, $\ket{\phi_n}=\ket{\psi_n}$ and $\hat\eta=\hat{\mathds{1}}$. Using Eqs. (\ref{eq:phi_eta_psi}, \ref{eq:biorthonormality}), it is immediate so see that a possible metric operator can be built from the eigenvectors of the hamiltonian as
\begin{eqnarray}\label{eq:eta_eigenvectors}
\hat\eta=\sum_{n}\ket{\phi_n}\bra{\phi_n}
\end{eqnarray}
and also
\begin{eqnarray}\label{eq:eta_inverse_eigenvectors}
\hat\eta^{-1}=\sum_{n}\ket{\psi_n}\bra{\psi_n}.
\end{eqnarray}
An important property of pseudo-hermitian operators in general is that their eigenvalues are either real or come in complex conjugate  pairs \cite{Bender_1998,Mostafazadeh_2002_431}. In particular, given the eigenvalue problem (\ref{eq:eigenvalue_eqn_H})
for an $\eta-$hermitian hamiltonian $\hat H$, it follows that
\begin{eqnarray}
    (E_i-E_j^*)\braket{\psi_i}{\psi_j}_\eta = 0.
\end{eqnarray}
Therefore, if
\begin{align}
        E_i^* \neq E_j \quad \text{then} \quad \braket{E_i}{E_j}_\eta = 0,
    \end{align}
so that orthogonality of states with different eigenvalues is guaranteed (in particular, but not exclusively, if such eigenvalues are real). Furthermore,
    \begin{equation}
        \forall \, E_i \in \mathbb{C} \quad \text{we have} \quad \lVert \ket{E_i}\rVert^2_\eta=\braket{E_i}{E_i}_\eta = 0,
    \end{equation}
that is, complex energy eigenvectors have zero norm, being thus not part of the physical states (in this regime, the metric is not positive-definite). By performing appropriate changes in the parameters of a pseudo-hermitian hamiltonian, one can often allow it to shift from a purely real spectrum to one with some complex values along with real ones (or even a completely complex spectrum).  The regime in which complex energies are present is called 
the ${\cal PT}$-broken phase, first described in \cite{Bender_1998}, and the boundary in parameters space at which the transition from real to complex spectrum takes place is an {\it exceptional surface} \cite{Kawabata2019,Jia:2023,Ryu:2024}.

We will consider more deeply the very interesting case of complex energies in an upcoming work \cite{work-in-progress}. In the present paper, we will work only with cases where the energy eigenvalues are real.
In our closing section, however, we will allow ourselves to make a conjecture about a possible interpretation of the ${\cal PT}-$broken phase in the context of High-Energy Physics.

Let us now consider the time evolution of a system with a pseudo-hermitian hamiltonian. Given the Schrödinger equation
\begin{eqnarray}
       \hat H \vert \psi \rangle = i \hbar \frac{d \, \vert \psi \rangle}{dt},
\end{eqnarray}
it follows that, for two arbitrary states $\ket{\psi}$ and $\ket{\phi}$,
\begin{align}
    \frac{d}{dt}{\langle \psi \vert \phi\rangle}_\eta 
    &=
    \frac{1}{i \hbar}
    \langle\psi \vert\left(\hat \eta \hat H - \hat H^\dagger\hat \eta \right)\vert \phi\rangle.
\end{align}
The above equation implies that the time invariance of the inner-product for any pair of states requires that the hamiltonian is \(\eta\)-hermitian, which is a more general condition than hermiticity. This condition was already pointed out by Pauli \cite{Pauli_1943} when commenting on the use of indefinite inner-products by Dirac \cite{Dirac_1942_180}. 

Another way to understand probability conservation in pseudo-hermitian closed systems is by discussing the time evolution of Heisenberg picture operators. Let us define a generic Heisenberg picture operator $\hat{\cal O}_H(t)$ such that it obeys the time evolution equation
\begin{eqnarray}\label{eq:Heisenberg_EOM}
    \frac{d\hat{\cal O}_H(t)}{dt} = \frac{1}{i\hbar}\left[\hat{\cal O}_H(t),\hat H\right],
\end{eqnarray}
where $\hat H$ is the (time-independent) system hamiltonian. The initial condition for the time evolution is taken to be the corresponding Schrödinger picture operator  $\hat{\cal O}_H(t=0)=\hat{\cal O}_S$. Equivalently, 
\begin{eqnarray}\label{eq:heisenberg-picture-operator}
    \hat{\cal O}_H(t) = \hat U(t)^{-1}\hat{\cal O}_S\hat U(t),
\end{eqnarray}
where $\hat U(t)=\exp\left(-i\hat H t/\hbar\right)$ is the time evolution operator. Note that, for a generic non-hermitian hamiltonian, $\hat H^\dagger\not=\hat H$, the evolution operator is not (Dirac-)unitary, $\hat U^\dagger\not=\hat U^{-1}$, so that probability conservation in the theory may be at risk. However, for a pseudo-hermitian hamiltonian, unitarity can be saved as long as the inner product used to calculate probabilities is the $\eta-$inner product defined in (\ref{eq:eta-inner-product}). Indeed, it is immediate to show that, for any $\eta-$hermitian hamiltonian, $\hat U^\# = \hat \eta \hat U^\dagger \hat \eta^{-1}=\hat U^{-1}$, and unitarity of time evolution is recovered, being equivalent to conservation of probability, as long as the $\eta-$inner product (\ref{eq:eta-inner-product}) is employed instead of the reference inner product \cite{Bender_2002_89,Bender:2007wu,Mannheim:2009zj}.

Observables must be such that not only their eigenvalues are real, but also that the  eigenvectors corresponding to different eigenvalues are mutually orthogonal. As such, they must be self-adjoint with respect to an adequate inner product. Of course, this property must stay true during time evolution. However, for a non-hermitian hamiltonian, hermiticity is in general broken under time evolution, i.e., for a given operator $\hat o(t_1)=\hat o^\dagger(t_1)$ at some time $t_1$, one has in general $\hat o(t_2)\not=\hat o^\dagger(t_2)$ for $t_2\not=t_1$. But given that the hamiltonian is $\eta-$hermitian, all other $\eta-$hermitian operators will keep this property under time evolution, i.e., if there is another time-dependent operator $\hat q(t)$ such that $\hat q(t_1)=\hat q^\#(t_1)$ at some time $t_1$, then $\hat q(t)=\hat q^\#(t)$ for all $t$, as long as $\hat H^\#=\hat H$.

Given the choice of the $\eta-$inner product (\ref{eq: eta-product}), observables $\hat O=\hat O^\#$ can be built from hermitian operators $\hat o=\hat o^\dagger$ as \cite{Mostafazadeh_2010_07}
\begin{eqnarray}
    \hat O = \hat\rho^{-1}\hat o\hat\rho,
\end{eqnarray}
where $\hat\rho=\hat\eta^{1/2}=\hat\rho^\dagger$. Note that the existence of a hermitian $\hat\rho$ requires a hermitian and positive metric operator $\hat \eta$. 

With the results above, we believe it should be clear that a proper treatment of any pseudo-hermitian system requires the knowledge of a nontrivial metric operator $\hat\eta$. With it, one can properly define the observables and establish the unitary time evolution of the theory in a consistent manner.

%%%%%%%%%%%%%%%%%%%%%%%%%%%%%%%%%%%%%%%%%%%%%%%%%%%%%%%%%%%%%%
\section{\label{sec:spectral-functions}On the Spectral functions of Pseudo-hermitian theories}
%%%%%%%%%%%%%%%%%%%%%%%%%%%%%%%%%%%%%%%%%%%%%%%%%%%%%%%%%%%%%%

Correlation functions are, of course, very relevant quantities in the study of any quantum theory. Indeed,  knowledge of the full set of  correlation functions of a quantum theory is equivalent to solving the theory. In the case of two-point correlation functions, the spectral function plays a central role, since it encodes much of the physical content of a theory \cite{Kallen:1952zz,Lehmann:1954_NuovoCimento,Peskin:1995ev}. 

Given the importance of the spectral function for the understanding of any theory, and also the interesting properties of the spectral function when the system hamiltonian is pseudo-hermitian, we decided to dedicate a whole section of this paper to it. We believe that the remarks we make here may have interesting consequences for the interpretation of quantum theories which display a spectral function which is not positive-definite, such as Yang-Mills theories \cite{1995IJMPA..10.1995O,1996CzJPh..46....1N,Alkofer:2000wg,2013MPLA...2830035C,Cucchieri_2005,Iritani_2009,Dudal_2014,Alkofer:2000wg,Alkofer_2004,Fischer:2020xnb,Kondo_2020,Cyrol_2016,Cyrol:2018_SciPost,Aguilar_2020,Dudal_2008,Vandersickel:2012tz} and Fermi liquids with multipolar interactions \cite{Aquino2020,Aquino2021,Aquino2024}.

Let us start by considering the generic time-ordered correlation function
\begin{eqnarray}\label{eq:generic_corr_funct}
    C_\eta^{AB}(t):={\langle \psi_0\vert \,T\bigl \{ \hat A(t) \hat B(0)\bigr\}\,\psi_0\rangle}_\eta,
\end{eqnarray}
where $T$ is the usual time ordering operator and $\ket{\psi_0}$ is the ground state. Note that we define the correlation function in terms of the $\eta-$inner product (\ref{eq:eta-inner-product}). Of course, if one would like to calculate the correlation function in terms of the reference (Dirac) inner product, it suffices to make the replacement $\hat\eta\rightarrow\hat{\mathds{1}}$. The correlation function in frequency space is given by the Fourier transform 
\begin{eqnarray}
    \tilde C_\eta^{AB}(\omega):=\int_{-\infty}^\infty dt\,e^{i\omega t}C_\eta^{AB}(t).
\end{eqnarray}
Inserting a complete set of energy eigenstates as in (\ref{eq:completeness}), we find, after a few steps, that
\onecolumngrid

\begin{eqnarray}
    \tilde{C}_\eta^{AB}(\omega)=\lim_{\epsilon\rightarrow0}\sum_m\frac{i}{2\pi}\left\{\
    \frac{\braket{\psi_0}{\hat A(0)\, m}_\eta \braket{m}{\hat B(0)\,\psi_0}_\eta}{\omega - \frac{E_m-E_0}{\hbar}+i\epsilon} -
    \frac{\braket{m}{\hat A(0)\,\psi_0}_\eta \braket{\psi_0}{\hat B(0)\,m}_\eta}{\omega + \frac{E_m-E_0}{\hbar}-i\epsilon}
    \right\}.
\end{eqnarray}
If $\hat A=\hat B$, then
\begin{eqnarray}
    \tilde{C}_\eta^{AA}(\omega)=\lim_{\epsilon\rightarrow0}\sum_m\frac{1}{2\pi}\frac{i}{\omega^2 - \left(\frac{E_m-E_0}{\hbar}\right)^2+i\epsilon}
    \frac{2(E_m-E_0)}{\hbar}
    \braket{\psi_0}{\hat A(0)\,m}_\eta \left(\braket{\psi_0}{\hat A^\#(0)\,m}_\eta\right)^*,
\end{eqnarray}
so that, for real eigenenergies, the correlation function can be written in the form of a K\"allen-Lehmann representation, i.e.,
\begin{eqnarray}\label{eq:Kallen-Lehmann-repr}
    \tilde{C}^{AA}_\eta(\omega) = \lim_{\epsilon\rightarrow0}\int_0^\infty ds\,\frac{i}{\omega^2-s+i\epsilon}\rho(s),
\end{eqnarray}
with
\begin{eqnarray}
    \rho^{AA}(s) = \sum_m\frac{1}{2\pi}
    \braopket{\psi_0}{\hat A(0)}{m}_\eta \left(\braket{\psi_0}{\hat A^\#(0)\,m}_\eta\right)^*
    \delta\left[ s - \left(\frac{E_m-E_0}{\hbar}\right)^2\right]
\end{eqnarray}
\twocolumngrid
being called the spectral function associated with the correlation function $C^{AA}_\eta$. Note that, in general, $\rho^{AA}(s)$ is a complex quantity. However, for $\hat A^\#=\hat A$, 
\begin{eqnarray}
    \rho^{AA}(s) = \lim_{\epsilon\rightarrow0}\sum_m\frac{1}{2\pi}
    \left|\braket{\psi_0}{\hat A(0)\,m}_\eta\right|^2
    \delta\left[ s - \left(\frac{E_m-E_0}{\hbar}\right)^2\right],
\end{eqnarray}
which is a non-negative quantity. Let us remark that, for a standard hermitian theory, $\hat\eta=\hat{\mathds{1}}$ the hermiticity condition corresponds to $\hat A^\dagger(t)=\hat A(t)$ and it also implies positivity of the spectral function, in this case.

In summary, we simply reviewed the well-known fact that the two-point correlation function of a self-adjoint operator has a positive spectral function. This has been done here simply to show explicitly that this result is also perfectly valid in the presence of a metric in Hilbert space, as long as the correlation functions properly take this metric into account. But mainly, we wish to stress an almost equivalent statement: the existence of a non-positive spectral function in some two-point correlation function can be evidence for a lack of \(\eta\)-hermiticity of the operators that define this correlation function. 

We believe that these fairly simple remarks can have interesting implications for the understanding of systems with exceptional points and/or pairs of complex conjugate energy eigenstates. We will comment further on this in Sec. \ref{sec:discussion}.

%%%%%%%%%%%%%%%%%%%%%%%%%%%%%%%%%%%%%%%%%%%%%%%%%%%%%%%%%%%%%%
\section{\label{sec:coupled-oscillators}A quantum mechanical toy model}
%%%%%%%%%%%%%%%%%%%%%%%%%%%%%%%%%%%%%%%%%%%%%%%%%%%%%%%%%%%%%%

%%%%
\subsection{Model hamiltonian, its associated metric operator, and equivalent hermitian hamiltonian}

Before proceeding to the more interesting case of quantum fields, let us first review the quantum mechanical case of two harmonic oscillators coupled by an imaginary bilinear term \cite{NANAYAKKARA200267,Fring_2018_51,Beygi_2015_91,Bender_2016_92,Felski_2018_coupled_osc_excited}, with a hamiltonian given by
\begin{equation}
    \label{eq:hamiltonian-coupled-osc}
    \hat H_{QM} ~=~ \frac{1}{2m} \left(\hat p_x^2 + \hat p_y^2\right) + \frac{m}{2} \left(\Omega_x^2 \hat x^2 + \Omega_y^2 \hat y^2\right) + ig\hat x \hat y~.
\end{equation}
Assuming that the position and momentum operators are hermitian,
\begin{eqnarray}
    \hat x^\dag &=& \hat x\nonumber\\
    \hat p_x^\dag &=& \hat p_x\nonumber\\
    \hat y^\dag &=& \hat y\nonumber\\
    \hat p_y^\dag &=& \hat p_y,
\end{eqnarray}
it is evident that (\ref{eq:hamiltonian-coupled-osc}) is not hermitian. Using the Baker-Haussdorff lemma \cite{Sakurai}, it is possible to show that \cite{Fring_2018_51}
\newcommand{\met}{e^{-2\theta \hat L_z/\hbar}}
\newcommand{\invmet}{e^{2\theta \hat L_z/\hbar}}
\begin{equation}
    \label{eq:pseudohermiticityH}
    \hat H_{QM} ^\dag = \met \hat H_{QM} \invmet 
\end{equation}
holds for 
\begin{equation}
    \label{eq:condition for the angle of the metric}
    \tanh 2\theta = \frac{2g}{m \left(\Omega_x^2 - \Omega_y^2\right)}.
\end{equation}
where $\hat L_z = \hat x\hat p_y - \hat y\hat p_x$. By inspecting (\ref{eq:pseudohermiticityH}), we identify the metric operator for this system as
\begin{eqnarray}
    \hat \eta = e^{-2\theta \hat L_z/\hbar} = \hat\eta^\dag,
\end{eqnarray}
Note that this form of the metric operator could be anticipated by considering that an added bilinear term $\lambda \hat x \hat y$ implements a rotation with respect to the axes of the independent position operators. In the case of a pair of harmonic oscillators, phase space trajectories of constant energy are ellipses in the XY plane. The distinguishing feature of our case is that the ``rotation'' in the ellipses is given by an imaginary angle.

It is evident that in order for the angular parameter \(\theta\) to be real we must have \(\lvert 2g \rvert < \lvert m \left(\Omega_y^2 - \Omega_x^2\right)\rvert\). We call this condition {\it weak coupling} and it is crucial for the hermiticity condition $\hat \eta^\dag=\hat \eta$. In this regime, we have a positive-defined metric operator \(\hat\eta = e^{-2\theta \hat L_z}\) and the hamiltonian \(\hat H_{QM}\) is said to be \(\eta\)-hermitian \cite{Mostafazadeh_2010_07}.

The usefulness of the metric operator can be appreciated in many aspects of the theory, as already discussed in Sec. \ref{sec:PT-symm-review}. A particularly interesting application is actually implemented through the definition of the operator $\hat \rho:=\hat\eta^{1/2}$, as it can perform a similarity transformation on the hamiltonian such that the operator
\begin{eqnarray}
    \hat h_{QM} := \hat \rho\hat H_{QM}\hat\rho^{-1}
\end{eqnarray}
is hermitian, i.e., $\hat h_{QM}^\dagger=\hat h_{QM}$. 
Whenever it can be defined, the spectrum of
$\hat h_{QM}$ is identical to that of $\hat H_{QM}$ \cite{Mostafazadeh_2010_07}. 
For this reason, $\hat h_{QM}$ is called the 
{\it equivalent hermitian hamiltonian} of 
$\hat H_{QM}$. For the system of oscillators is an 
imaginary coupling (\ref{eq:hamiltonian-coupled-osc}), 
\begin{align}
    \label{eq: Equivalent hamiltonian}
    \hat h_{QM} &= e^{-\theta \hat L_z/\hbar} \hat H_{QM} e^{\theta \hat L_z/\hbar} \nonumber\\
    &=
    \frac{1}{2m} \left(\hat p_x^2 + \hat p_y^2\right) + \frac{m}{2} \left(\omega_x^2 \hat x^2 + \omega_y^2 \hat y^2\right) ,
\end{align}
where
\begin{eqnarray}
    \label{eq:omegax_omegay}
    \omega_x^2&:=&\frac{\cosh^2\theta ~\Omega_x^2 + \sinh^2 \theta~\Omega_y^2}{\cosh2\theta} \nonumber\\
  &=&\frac12\left[
  \Omega_x^2+\Omega_y^2-\sqrt{(\Omega_x^2-\Omega_y^2)^2-4g^2/m^2}\right],\nonumber\\  
    \omega_y^2 &:=&\frac{\sinh^2\theta~ \Omega_x^2 + \cosh ^2 \theta ~\Omega_y^2}{\cosh 2\theta}\nonumber\\
    &=&\frac12\left[
\Omega_x^2+\Omega_y^2+\sqrt{(\Omega_x^2-\Omega_y^2)^2-4g^2/m^2}\right],
\end{eqnarray}
where we used (\ref{eq:condition for the angle of the metric}). Notice that the hermiticity of the $\hat\rho$ operator is crucial for the existence of the hermitian hamiltonian $\hat h_{QM}$. In the case of the hamiltonian (\ref{eq:hamiltonian-coupled-osc}), this is only possible in the weak coupling regime, where $\theta\in\mathbb{R}$.

Since the spectra of $\hat H_{QM}$ 
and of $\hat h_{QM}$ are identical, it is immediate to see that the energy eigenvalues of the hamiltonian (\ref{eq:hamiltonian-coupled-osc}) are 
\begin{align}\label{eq:HO_spectrum}
    e_{QM}(n_x,n_y) &= E_{QM}(n_x,n_y) \hspace{1cm} n_x, n_y \in \IN, \nonumber\\
    &= \hbar \omega_x \left(n_x+\frac{1}{2}\right) + \hbar \omega_y \left ( n_y + \frac{1}{2}\right), 
\end{align}
where \(e_{QM}\) denotes the eigenvalues of \(\hat h_{QM}\) and \(E_{QM}\) denotes the eigenvalues of \(\hat H_{QM}\). 

The energy eigenvectors $\ket{\psi_n}$ of $\hat H_{QM}$ can be calculated from those of $\hat h_{QM}$, $\ket{\chi}$, as \cite{Mostafazadeh_2010_07}
\begin{eqnarray}
    \ket{\psi_n} = \hat\rho^{-1}\ket{\chi}.
\end{eqnarray}

We have just calculated the spectrum of the non-hermitian hamiltonian (\ref{eq:hamiltonian-coupled-osc}) and showed that it is real and bounded from below. To achieve that goal, we resorted to the equivalent hamiltonian (\ref{eq: Equivalent hamiltonian}). However, this is not the only possible method. In the next subsection, we show a direct calculation of the spectrum of the hamiltonian and the corresponding energy eigenstates using a direct algebraic method. Whenever the spectrum is real, for a positive metric, the equivalent hamiltonian exists and both  methods are applicable.

%%%%%%%%%%%%%%%%%
\subsection{Observables and ladder operators}

In the previous subsection we used the equivalent hamiltonian to solve the eigenvalue problem for our original hamiltonian because the Dirac inner product and the standard harmonic oscillator are objects we are more familiar with. This procedure can be compared to an ``active transformation'' effected by the square-root of the metric, from which the new hamiltonian (\ref{eq: Equivalent hamiltonian}) results. In this subsection, we will adopt a sort of ``passive transformation'' point of view, in which we define new variables and tackle the same problem directly, i.e., in terms of the original hamiltonian (\ref{eq:hamiltonian-coupled-osc}).

Let us define the new variables
\begin{align}
    \label{eq: X, Y}
    \hat X &= \hat\rho^{-1}\hat x\hat\rho=\hat x \cosh \theta + i \hat y \sinh \theta \nonumber \\
    \hat Y &= \hat\rho^{-1}\hat y\hat\rho=-i\hat x \sinh \theta + \hat y \cosh \theta\nonumber
    \\
    \hat P_X &= \hat\rho^{-1}\hat p_x\hat\rho=\hat p_x \cosh \theta + i \hat p_y \sinh \theta 
    \nonumber \\
    \hat P_Y &= \hat\rho^{-1}\hat p_y\hat\rho=-i\hat p_x \sinh \theta + \hat p_y \cosh \theta.
\end{align}

Note that these variables can be formally interpreted as the result of a sort of ``rotation'' in the reference frame of position variables by an imaginary angle $i\theta$. They obey the familiar canonical commutation relations
\begin{align}
    \label{eq: Commutation for the new position and momentum operators}
    \left[\hat X, \hat P_j \right] &= i \hbar \delta_{jX} \hat{\mathds{1}}
    \nonumber\\
    \left[\hat Y, \hat P_j \right] &= i \hbar \delta_{jY} \hat{\mathds{1}}
    \nonumber\\
    \left[ \hat X, \hat Y\right ] &= 0
    \nonumber \\
    \left[\hat P_X, \hat P_Y \right] &= 0
\end{align}
where \(j= X,Y\).

It is straightforward to show that these variables are $\eta-$hermitian, i.e.,
\begin{eqnarray}\label{eq:eta-hermiticity-of-QM-observables}
    \hat X^\# &=& \hat X\nonumber\\
    \hat Y^\# &=& \hat Y\nonumber\\
    \hat P_X^\# &=& \hat P_X\nonumber\\
    \hat P_Y^\# &=& \hat P_Y
\end{eqnarray}
and are identified as observables of the system \cite{Mostafazadeh_2010_07}. Indeed, since the $\eta-$inner product (\ref{eq:eta-inner-product}) is an adequate inner product for the quantum system defined by the hamiltonian (\ref{eq:hamiltonian-coupled-osc}), the eigenvalues of the operators (\ref{eq: X, Y}) are all real and their eigenvectors are orthogonal with respect to the $\eta-$inner product.

With these new variables, we may write the hamiltonian \(\hat H_{QM}\) as
\begin{equation}
    \label{eq: Hamiltonian with nem position and momentum operators}
    \hat H_{QM} = \frac{1}{2m} \left(\hat P_X^2 + \hat P_Y^2\right) + \frac{m}{2} \left(\omega_x^2 \hat X^2 + \omega_y^2 \hat Y^2 \right).
\end{equation}
Note that this hamiltonian is precisely the same as (\ref{eq:hamiltonian-coupled-osc}), but now written in terms of the new variables in (\ref{eq: X, Y}). Indeed, since the dynamical variables in (\ref{eq: Hamiltonian with nem position and momentum operators}) are not hermitian, but rather $\eta-$hermitian, it is evident that $\hat H_{QM}^\dagger\not=\hat H_{QM}$ and $\hat H_{QM}^\#=\hat H_{QM}$.

Of course, given the form of the hamiltonian in (\ref{eq: Hamiltonian with nem position and momentum operators}), its quantization in the Hilbert space endowed with the $\eta-$inner product is trivial. However, for the sake of making the relationship between both Hilbert spaces (the one with the reference (Dirac) inner product and the one with the $\eta-$inner product) as clear as possible, let us insist in the reference space for a while.

Now, note that we have not changed spaces since \(\hat H_{QM}\) is not (Dirac) hermitian, meaning the ''passive rotation`` does not affect the Hilbert space being considered as was the case for the equivalent Hamiltonian (\ref{eq: Equivalent hamiltonian}). Mirroring the procedure for solving the standard harmonic oscillator, we examine the possibility of establishing ladder operators in the new Hilbert space. With this in mind, we define
\begin{align}\label{eq:HO-ladder operators}
    \hat \alpha_{X}^\dag &\coloneq 
    \sqrt{\frac{m\omega_{x}}{2 \hbar}}
    \left (
        \hat X -\frac{i}{m\omega_x} \hat P_X
    \right)
    \nonumber \\
    \hat \beta_X &\coloneq 
    \sqrt{\frac{m\omega_x}{2\hbar}}
    \left (
        \hat X + \frac{i}{m \omega_x}\hat P_X
    \right )
    \nonumber\\
    \hat \alpha_{Y}^\dag &\coloneq 
    \sqrt{\frac{m\omega_{y}}{2 \hbar}}
    \left (
        \hat Y -\frac{i}{m\omega_y} \hat P_Y
    \right)
    \nonumber\\
    \hat \beta_Y &\coloneq 
    \sqrt{\frac{m\omega_y}{2\hbar}}
    \left (
        \hat Y + \frac{i}{m \omega_y}\hat P_Y
    \right ).
\end{align}

Since the familiar commutation relations apply in regards to the previously defined operators \(\hat X\), \(\hat Y\), \(\hat P_X\), \(\hat P_Y\), it is then easy to show that the hamiltonian \(\hat H_{QM}\) will possess the familiar (but still non-hermitian) structure
\begin{align}
    \hat H_{QM} = \hbar \omega_x \left ( \hat \alpha_X^\dag \hat \beta_X + \frac{1}{2}\right) 
    +
    \hbar \omega_y \left( \hat \alpha_Y^\dag \hat \beta_Y + \frac{1}{2}\right).
\end{align}

The distinguishing feature of the above hamiltonian is that the operators being multiplied are not Dirac conjugates of one another (i.e., $\hat \alpha_X^\dagger\not=\hat\beta_X^\dagger$ etc.). In fact, these operators are related by
\begin{equation}
    \label{eq: eta pseudo hermitian of b dag is a}
    \hat \alpha_i^\dag = \hat\eta^{-1}\hat \beta_i^\dag\hat\eta = \hat\beta_i^\#,
\end{equation}
where $i=X,Y$. This shows us that, since the $\eta-$norm of any state $\hat\beta_i\ket{\psi}$ is non-negative, i.e.,
\begin{align}
    {\langle \hat \beta_i \psi \vert \hat \beta_i \psi\rangle}_{\eta} 
    &=
    \langle \psi \vert \hat \beta_i^\dag \hat \eta \hat \beta_i \vert \psi \rangle  \nonumber\\
    &=
    \langle \psi \vert  \hat \eta \hat \alpha_i^\dag \hat \beta_i \vert \psi \rangle  \nonumber\\
    &=
    {\langle \psi \vert \hat \alpha_i^\dag\hat \beta_i \psi\rangle}_{\eta} \geq 0,
\end{align}
then the expectation value of the hamiltonian (\ref{eq:hamiltonian-coupled-osc}) is positive-defined. Finally, given that
\begin{align}
    \label{eq: New ladder ops commutation relations}
    &\left[\hat \beta_i,\hat \alpha_j^\dag \right] = \hat{\mathds{1}} \delta_{ij} \nonumber\\
    &\left[ \hat H_{QM}, \hat \beta_i \right]= -\hbar \omega_i\hat \beta_i
    \nonumber \\
    &\left[ \hat H_{QM}, \hat \alpha_i^\dag \right]= \hbar \omega_i\hat \alpha_i^\dag,
\end{align}
(where no summation over repeated indices is implied) it is straightforward to recover the result (\ref{eq:HO_spectrum}) for the energy spectrum. 

As for the usual 2-d harmonic oscillator, the ground state $\ket{\psi_0}$ is defined as 
\begin{eqnarray}\label{eq:HO-ground-state}
    \hat\beta_X \ket{\psi_0}=0\;\;\;\;\;\;\mbox{and}\;\;\;\;\;\;\hat\beta_Y\ket{\psi_0}=0,
\end{eqnarray}
whereas the stationary excited states can be obtained from the ground state by repeated application of the raising operators, i.e.,
\begin{eqnarray}\label{eq:HO-excited-states}
    \ket{N_X,N_Y} &=& \frac{1}{\sqrt{N_X!N_Y!}}\left(\hat \alpha_X^\dagger\right)^{N_X}\left(\hat \alpha_Y^\dagger\right)^{N_Y}\ket{\psi_0}\nonumber\\
     &=&\frac{1}{\sqrt{N_X!N_Y!}}\left(\hat \beta_X^\#\right)^{N_X}\left(\hat \beta_Y^\#\right)^{N_Y}\ket{\psi_0}.
\end{eqnarray}
The result we arrived at above is exactly the same as for the equivalent hamiltonian, as had to be the case. Therefore, we have shown that we are able to perform the same analysis of the energy eigenvalues directly from \(\hat H_{QM}\). For this, the usage of the inner product 
\({\langle\cdot\vert\cdot\rangle}_\eta\) was essential. In the process we have identified ladder operators in this space, namely we identify \(\hat \beta_i\) as the lowering operators and \(\hat\beta_i^\#=\hat \alpha_i^\dagger\) as the raising operators. %Once again we call the reader's attention to the fact that both operators are not the Dirac transposed of one another (in the reference Hilbert space representation). Instead, as we verified, \(\hat \alpha_i^\dag\) is the \(\eta\)-hermitian adjoint of \(\hat \beta_i\), as stated in (\ref{eq: eta pseudo hermitian of b dag is a}).

%%%%%%%%%%%%%%%%%%%%%%%%%%%%%%%%%%%%%%%%%%%
\subsection{Time evolution of Heisenberg picture operators and unitarity}

In what follows, we will be interested in the time evolution of dynamical variables of the system, i.e., of Heisenberg picture operators $\hat{\cal O}_H$ such that their Schr\"odinger picture counterpart is $\hat{\cal O}_S \in \{\hat x, \hat X, \hat y,\hat Y,\hat p_x,\hat P_X,\hat p_y,\hat P_Y\}$, previously defined in the text. 

Let us start by the Heisenberg picture operators corresponding to the $\eta-$hermitian operators in Eq. (\ref{eq: X, Y}). Since the hamiltonian written in terms of these dynamical variables (\ref{eq: Hamiltonian with nem position and momentum operators}) is a sum of two decoupled parts, it is easy to see that
\begin{eqnarray}\label{eq:time-evolution-observables}
    \hat X_H(t) &=& \cos(\omega_xt)\hat X+\sin(\omega_xt)\frac{\hat{P}_X}{m\omega_x}\nonumber\\
    \hat Y_H(t) &=& \cos(\omega_yt)\hat Y+\sin(\omega_yt)\frac{\hat{P}_Y}{m\omega_y}\nonumber\\
    \hat P_{XH}(t) &=& \cos(\omega_x t)\hat P_X -m\omega_x\sin(\omega_xt)\hat X\nonumber\\
    \hat P_{YH}(t) &=& \cos(\omega_y t)\hat P_Y -m\omega_y\sin(\omega_yt)\hat Y ,
\end{eqnarray}
where the Schr\"odinger picture operators $\hat X$, $\hat Y$, $\hat P_X$, and $\hat P_Y$ are defined in (\ref{eq: X, Y}). The Heisenberg picture operators in (\ref{eq:time-evolution-observables}) can  be obtained either by solving the Heisenberg equation of motion (\ref{eq:Heisenberg_EOM}), or by application of the time evolution operator $\hat U(t)=\exp\left(-i\hat H_{QM} t\right)$, as in (\ref{eq:heisenberg-picture-operator}). Since $\hat H_{QM}^\#=\hat H_{QM}$, it immediately follows that the time evolution operator is unitary with respect to the $\eta-$inner product, that is, $\hat U^\#(t)=\hat U^{-1}(t)$.

One can also quickly check that the operators in (\ref{eq:time-evolution-observables}) are $\eta-$hermitian for any time $t$: $\hat X_H(t)^\#=\hat X_H(t)$, $Y_H(t)^\#=\hat Y_H(t)$ etc. This property, which can be derived from the $\eta-$hermiticity of the hamiltonian, is of course fully compatible with such operators being associated with physical observables.

For comparison, let us also consider the time evolution of the original variables  of the Hamiltonian (\ref{eq:hamiltonian-coupled-osc}). The solution to the Heisenberg picture equation of motion (\ref{eq:Heisenberg_EOM}), with the initial conditions $\hat x_H(t=0)=\hat x$, $\hat y_H(t=0)=\hat y$, $\hat p_{xH}(t=0)=\hat p_x$, and $\hat p_{yH}(t=0)=\hat p_y$ can be easily be found from (\ref{eq: X, Y}) and (\ref{eq:time-evolution-observables}) and is given by 
\begin{eqnarray}\label{eq:time-evolution-original-variables}
    \hat x_H(t) &=& \cosh\theta\hat X_H(t)-i\sinh\theta\hat Y_H(t)\nonumber\\
    \hat y_H(t) &=& i\sinh\theta\hat X_H(t) + \cosh\theta\hat Y_H(t)\nonumber\\
    \hat p_{xH}(t) &=& \cosh\theta\hat P_{XH}(t)-i\sinh\theta\hat P_{YH}(t)\nonumber\\
    \hat p_{yH}(t) &=&  i\sinh\theta\hat P_{XH}(t) + \cosh\theta\hat P_{YH}(t).
\end{eqnarray}

It is interesting to note that, although $\hat x^\dagger=\hat x$, this hermiticity condition is not valid in general for $t\not=0$, i.e., $\hat x_H^\dagger(t)\not=\hat x_H(t)$. Evidently, this is due to lack of hermiticity of the hamiltonian itself. Let us also recall that this operator is also not $\eta-$hermitian, i.e., $\hat x_H(t)^\#\not=\hat x_H(t)$. Such a property is another way of understanding that the $\hat x$ operator cannot be associated with an observable of the system. 

Let us conclude this subsection by remarking that the properties of the $\eta-$hermitian operators discussed so far are evidence that not only the $\eta-$inner product is more adequate than the reference inner product for dealing with the hamiltonian (\ref{eq:hamiltonian-coupled-osc}) (or, equivalently, \ref{eq: Hamiltonian with nem position and momentum operators}), but also the $\eta-$hermitian operators make the physics more apparent
than the (Dirac) hermitian operators used in the original hamiltonian (\ref{eq:hamiltonian-coupled-osc}). In the next subsection, we will show how one can arrive at a similar conclusion from the point of view of correlation functions of products of operators.

%%%%%%%%%%%%%%%%%%%%%%%%%%%%%%%%%%%%%%%%%%%
\subsection{Correlation functions and spectral function}

In order to complete our discussion on the quantum mechanical case, let us now discuss a few correlation functions. Two crucial aspects of the following discussion are, first, the adequate choice of operators involved in the correlation function and, second, the choice of inner product (since every correlation function is calculated assuming some inner product). 

The first (and simplest) correlation function we consider is the ground-state expectation value \footnote{From this section onwards, we shall drop the $H$ subscript for Heisenberg picture operators, whenever this brings no confusion to the reader.}
\onecolumngrid
\begin{equation}
    \label{eq:XX-propagator}
   C^{XX}_\eta(t):= {\langle \psi_0\vert \,T\bigl \{ \hat X(t) \hat X(0)\bigr\}\,\psi_0\rangle}_\eta 
    =\frac{\hbar}{2m\omega_x}\left[
    \theta(t) e^{-i\omega_xt} 
    +
    \theta (-t) e^{i\omega_x t}\right]
   % =
    %\lim_{\epsilon\to0} \, \, \frac{\hbar}{m}\int_{-\infty}^{\infty}\frac{dk}{2\pi}\frac{i}{k^2-\omega_x^2 + i\epsilon}\,e^{ikt},
\end{equation}
where we used (\ref{eq:HO-ladder operators}) and (\ref{eq:time-evolution-observables}), and $T$ is the usual time-ordering operator. Given that the model hamiltonian (\ref{eq:hamiltonian-coupled-osc}) is equal to the $\eta-$hermitian Eq. (\ref{eq: Hamiltonian with nem position and momentum operators}), written in terms of operators such as $\hat X$, the result (\ref{eq:XX-propagator}) could be easily anticipated. Finally, note that the correlation function (\ref{eq:XX-propagator}) is defined in terms of the $\eta-$inner product. In other words, the operators whose product is calculated (in this case, $\hat X$) are self-adjoint with respect to the inner product used in the definition of this correlation function. For the case of the coordinate $\hat Y$, a completely analogous result follows
\begin{equation}
    \label{eq:YY-propagator}
    C^{YY}_\eta(t):={\langle \psi_0\vert \,T\bigl \{ \hat Y(t) \hat Y(0)\bigr\}\,\psi_0\rangle}_\eta 
    =
\frac{\hbar}{2m\omega_y}\left[
    \theta(t) e^{-i\omega_yt} 
    +
    \theta (-t) e^{i\omega_y t}\right].
\end{equation}
An important quantity that can be obtained from the 
correlation functions above is the spectral function 
\cite{Peskin:1995ev}. The Fourier transform of the
two-point function (\ref{eq:XX-propagator}) is given by
\begin{eqnarray}
    \tilde{C}^{XX}_\eta(\omega)&:=&\int_{-\infty}^\infty\,dt\,e^{i\omega t}C^{XX}_\eta(t)\nonumber\\
    &=& \lim_{\epsilon\rightarrow0^+}\frac{\hbar}{m}\frac{i}{\omega^2-\omega_x^2+i\epsilon}
\end{eqnarray}
which can be trivially put in the K\"all\'en-Lehmann representation
\begin{eqnarray}
   \tilde{C}^{XX}_\eta(\omega)&=& \lim_{\epsilon\rightarrow0^+}\int_0^\infty\frac{ds}{2\pi}\frac{i}{\omega^2-s+i\epsilon}\rho(s),
\end{eqnarray}
where
\begin{eqnarray}
    \rho^{XX}(s)=\frac{2\pi\hbar}{m}
    \delta(s-\omega_x^2)\geq 0,\;\;\;\;\;\forall s
\end{eqnarray}
is the spectral function. Note that it is non-negative 
everywhere. This could be anticipated on general 
grounds, since $\hat X(t)$ is self-adjoint with respect 
to the inner product used to define the correlation 
function (\ref{eq:XX-propagator}).

It is now interesting to calculate the correlation function with one of the dynamical variables of the ``original'' pseudo-hermitian hamiltonian (\ref{eq:hamiltonian-coupled-osc}). From (\ref{eq:time-evolution-original-variables}),
we have the correlation function
\begin{align}\label{eq:c_eta_xx}
    C_\eta^{xx}(t):={\langle \psi_0 \vert \,T\bigl\{\hat x(t) \hat x (0)\bigr\}\,\psi_0 \rangle}_\eta 
    &=
    \cosh^2 \theta \:{\langle \psi_0\vert \,T\bigl \{ \hat X(t) \hat X(0)\bigr\}\,\psi_0\rangle}_\eta
    -
    \sinh^2\theta  \:{\langle \psi_0\vert \,T\bigl \{ \hat Y(t) \hat Y(0)\bigr\}\,\psi_0\rangle}_\eta,
%    &=
%    \lim_{\epsilon \,\to \,0}\; \frac{\hbar}{m}
%    \left( 
%        \cosh^2\theta \int_{-\infty}^{\infty} \frac{dk}{2\pi}\frac{i}{k^2-\omega_x^2 +i\epsilon}\:e^{ikt}
 %       -
 %       \sinh^2 \theta \int_{-\infty}^{\infty} \frac{dl}{2\pi}\frac{i}{l^2-\omega_y^2 +i\epsilon}\:e^{ilt}
 %   \right)
\end{align}
whose Fourier transform is given by
\begin{eqnarray}
    \tilde{C}_\eta^{xx}(\omega)=\lim_{\epsilon\rightarrow0^+}\frac{\hbar}{m}\left[
    \frac{i}{\omega^2-\omega_x^2+i\epsilon}\cosh^2\theta-\frac{i}{\omega^2-\omega_y^2+i\epsilon}\sinh^2\theta
    \right].
\end{eqnarray}
\twocolumngrid
If we try to express $\tilde{C}_\eta^{xx}(\omega)$ 
in a way analogous to the K\"allen-Lehmann representation, we find a would-be spectral function
\begin{eqnarray}\label{eq:rho_xx}
    \rho^{xx}(s)= \frac{\pi \hbar}{m} 
    \left[
      \delta\left(s -\omega_x^2\right) \, \cosh ^2 \theta 
        -
        \delta\left(s -\omega_y^2\right)
        \sinh^2 \theta 
    \right],
\end{eqnarray}
which is {\it not} a positive-defined function everywhere. The correlation function $C_\eta^{yy}$ is very similar to (\ref{eq:c_eta_xx}) and its associated spectral function $\rho^{yy}$ is also not everywhere positive. This property is often called {\it positivity violation}. Indeed, positivity of the spectral function would be expected from two-point functions of hermitian operators. In the case of the operator $\hat x(t)$, its hermiticity is valid only at $t=0$, due to the nonhermitian nature of the hamiltonian (\ref{eq:hamiltonian-coupled-osc}) itself. Therefore, it is not surprising that its two-point correlation function violates reflection positivity.

In this specific example, we saw that positivity violation in a correlation function involving only hermitian operators is a consequence of the non-hermiticity of the hamiltonian. We believe that this can be a generic feature of quantum theories.

%%%%%%%%%%%%%%%%%%%%%%%%%%%%%%%%%%%%%%%%%%%%%%%%%%%%%%%%%%%%%%
\section{A scalar field model with imaginary coupling}
\label{sec:qft-model} 
%%%%%%%%%%%%%%%%%%%%%%%%%%%%%%%%%%%%%%%%%%%%%%%%%%%%%%%%%%%%%%

%\cite{bender2023ptsymmetric}

Having discussed the quantum mechanical model of 
a pair of harmonic oscillators with an imaginary coupling, 
we are now in a good position to study a quantum field theory 
that is a natural generalization of such a model: a pair of scalar 
fields with an imaginary bilinear coupling. But before looking at 
the quantum theory, let us briefly investigate the classical 
field-theoretical model and later on we shall proceed to its 
canonical quantization.

%%%
\subsection{The model lagrangian and conserved quantities}

As our starting point, let us consider the action of the theory in natural units $(\hbar=c=1)$
\onecolumngrid
\begin{align}\label{eq:model_lagrangian}
    S = \int d^4x{\cal L} 
    = \int d^4x\left[\frac{1}{2}\partial^\mu a\partial_\mu a - \frac{m_a^2}{2}a^2 + \frac{1}{2}\partial^\mu \varphi\partial_\mu \varphi - \frac{m_\varphi^2}{2}\varphi^2 - iga\varphi\right] ,
\end{align}
\twocolumngrid
where ${\cal L}$ is the lagrangian density, $a=a(x^\mu)$ and $\varphi=\varphi(x^\mu)$ are scalar fields, $m_a$ and $m_\varphi$ are massive parameters of the model and $g\in\mathbb{R}$ is the coupling parameter. Note that, for the case of $d=1$ spacetime dimensions (i.e., only a time dimension), the theory corresponds precisely to the hamiltonian we have studied in the previous section.

The lagrangian density (\ref{eq:model_lagrangian}) is evidently complex. The corresponding Euler-Lagrange equations are
\begin{eqnarray}\label{eq:eom_A_phi}
    (\Box+m_a^2)a &=& -ig\varphi\nonumber\\
    (\Box+m_\varphi^2)\varphi &=& -iga,
\end{eqnarray}
where $\Box=\partial^2/\partial t^2 - \nabla^2$ is the D'Alembertian operator. 

Note that, even if the initial and/or boundary conditions are such that the fields $a$ and $\varphi$ are real, the imaginary coupling will necessarily ``push'' the $a$ and the $\varphi$ fields into the complex plane. In other words, for any nontrivial initial condition, the classical $a$ and $\varphi$ fields will not remain real under time evolution. 

At the quantum level, this same property will be manifest as follows: if, for a given choice of Hilbert space inner product the (Heisenberg picture) field operators are hermitian at a given time $t_0$, then the same field operators at some other time $t_1\not=t_0$ will be, in the general case, non-hermitian. This is typical of pseudo-hermitian systems and is a direct consequence of the non-observability of the field operators for the theory defined by (\ref{eq:model_lagrangian}). However, this should not be viewed as a fundamental problem with the theory. Rather, it only means that the dynamical variables used to write down the action (\ref{eq:model_lagrangian}) do not correspond to observables. As in ordinary Quantum Mechanics, this property does not imply that the theory itself lacks observables. Indeed, observables can be built from the (non-selfadjoint) dynamical variables, as we shall discuss later.

%%%%%%%%%%%

In order to find a hamiltonian for the system, let us 
first derive the energy-momentum tensor. 
%But before proceeding, let us remark that this may be a nontrivial task in pseudo-hermitian theories. As discussed in \cite{Alexandre:2017foi}, Noether's theorem may not apply to some pseudo-hermitian theories, i.e., conserved currents may have no associate symmetries and, conversely, some symmetries could not imply the existence of some conserved current. In the case of the action (\ref{eq:model_lagrangian}), however, it is straightforward to show that the Poincar\'e invariance of the action does lead to a conserved energy-momentum tensor, as implied by Noether's theorem. 
%Indeed, 
Just like in an ordinary, real, theory we find
\begin{eqnarray}
    T^{\mu\nu} &=& \frac{\partial{\cal L}}{\partial(\partial_\mu\varphi)}\partial^\nu\varphi + \frac{\partial{\cal L}}{\partial(\partial_\mu A)}\partial^\nu a - {\cal L}\eta^{\mu\nu}
\end{eqnarray}
where $\eta_{\mu\nu}={\rm diag}(1,-1,-1,-1)$ are the components of the Minkowski space metric tensor.
It should not be surprising that, from a 
classical standpoint, the energy-momentum tensor 
associated with the action 
(\ref{eq:model_lagrangian}) is a complex quantity.

The hamiltonian is then given by
\onecolumngrid
\begin{eqnarray}\label{eq:field_hamiltonian}
    H &=& \int d^3x\,T^{00} %\nonumber\\    &=& 
    = \int d^3x\,\left[\frac{1}{2}\pi_\varphi^2 + \frac12\left(\nabla\varphi\right)^2 + \frac{m_\varphi^2}{2}\varphi^2 + \frac{1}{2}\pi_a^2 + \frac12\left(\nabla a\right)^2 + \frac{m_a^2}{2}a^2 + iga\varphi\right],
\end{eqnarray}
\twocolumngrid
where the canonically conjugate field momenta are
\begin{eqnarray}
    \pi_\varphi &=& \frac{\partial{\cal L}}{\partial(\partial^0\varphi)} = \frac{\partial\varphi}{\partial t},\nonumber\\
    \pi_a &=& \frac{\partial{\cal L}}{\partial(\partial^0a)} = \frac{\partial a}{\partial t}.
\end{eqnarray}

The linear momentum carried by the fields is
\begin{eqnarray}\label{eq:linear_momentum}
    \vec {\cal P} 
    &=& \int d^3x\,T^{0i}\hat{e}_i= \int d^3x\,\left[\Pi_\varphi\vec\nabla\varphi + \Pi_a\vec\nabla a\right],
\end{eqnarray}
where $\hat{e}_i$ ($i=1,2,3$) are the usual cartesian unit vectors.

Note that, although the functional form of $\vec{\cal P}$ is not complex, the time evolution of the classical fields will typically lead to complex values of linear momentum, since the fields $\varphi$ and $a$ become complex themselves.

All these features of the classical theory could make it seem a 
meaningless task to define a quantum theory having the complex 
action (\ref{eq:model_lagrangian}) as a starting 
point. However, as we will argue next, it is actually quite simple
to build a meaningful quantum field theory from it.

%%%%%%%%%%
\subsection{Canonical quantization}

Following the standard procedure of canonical 
quantization, let us promote the dynamical 
variables to (Schrödinger picture) operators, such that they respect 
the standard equal-time commutation relations
\begin{eqnarray}
    [\hat\varphi(\vec x),\hat\varphi(\vec{y})]=
    [\hat a(\vec x),\hat a(\vec{y})]&=&0,\nonumber\\
\left[\hat \varphi(\vec x),\hat \pi_\varphi(\vec y)\right]=
\left[\hat a(\vec{x}),\hat \pi_a(\vec{y})\right]&=& i\delta^{(3)}(\vec{x}-\vec y)\mathds{\hat1}.
\end{eqnarray}

If one assumes that the fields and their conjugate momenta are hermitian with respect to some (reference) inner product, i.e.,
\begin{eqnarray}
    [\hat \varphi(\vec x)]^\dagger &=& \hat \varphi(\vec x)\nonumber\\
    \left[\hat a(\vec x)\right]^\dagger &=& \hat a(\vec x),
\end{eqnarray}
then the hamiltonian (\ref{eq:field_hamiltonian}) 
is, of course, not hermitian due to the 
imaginary coupling term $ig\hat a\hat \varphi$. 

In order to establish the standard analogy to the case of coupled harmonic oscillators
discussed in Sec. \ref{sec:coupled-oscillators}, let us first define 
the Fourier-transformed fields and momenta
\begin{eqnarray}
    \tilde\psi(\vec p) = \int d^3x\,e^{i\vec p\cdot\vec x}\psi(\vec x)
\end{eqnarray}
where $\psi = \varphi,a,\pi_\varphi,$ or $\pi_A$. From now on, we shall omit the caret notation ($\hat{\,}$) to denote operators, except to avoid any possible confusion. The commutation relations between the Fourier-transformed fields are
\begin{align}
    &[\tilde\varphi(\vec p),\tilde\varphi(\vec{k})]=
    [\tilde a(\vec p),\tilde a(\vec{k})]=[\tilde \varphi(\vec p),\tilde a(\vec{k})]=0\nonumber\\
  &\left[\tilde \pi_\varphi(\vec p),\tilde \pi_\varphi(\vec k)\right]=\left[\tilde \pi_a(\vec p),\tilde \pi_a(\vec k)\right] =\left[\tilde \pi_\varphi(\vec p),\tilde \pi_a(\vec k)\right]=0\nonumber\\
  &\left[\tilde \varphi(\vec p),\tilde \pi_a(\vec k)\right]=\left[\tilde a(\vec p),\tilde \pi_\varphi(\vec k)\right]=0\nonumber\\
  &\left[\tilde \varphi(\vec p),\tilde \pi_\varphi(\vec k)\right]=
  \left[\tilde a(\vec{p}),\tilde \pi_a(\vec{k})\right]= i(2\pi)^3\delta^{(3)}(\vec{p}+\vec k).
\end{align}

In terms of these operators, 
the (Schrödinger picture) canonically quantized version of the hamiltonian (\ref{eq:field_hamiltonian}) reads
\onecolumngrid
\begin{eqnarray}\label{eq:field_hamiltonian_fourier_coupled}
    H = \frac12\int \frac{d^3p}{(2\pi)^3}\left[
    \tilde \pi_\varphi(-\vec p)\tilde \pi_\varphi(\vec p) +
    \omega_\varphi^2(\vec p)\tilde\varphi(-\vec p)\tilde\varphi(\vec p) +
    \tilde \pi_a(-\vec p)\tilde \pi_a(\vec p) +
    \omega_a^2(\vec p)\tilde a(-\vec p)\tilde a(\vec p) +
      2ig\tilde a(-\vec p)\tilde\varphi(\vec p)
    \right],
\end{eqnarray}
\twocolumngrid
where $\omega_\varphi^2(\vec p):=\vec p^2+m_\varphi^2$ and 
$\omega_a^2(\vec p):=\vec p^2+m_a^2$. It is not surprising that our hamiltonian is simply a (continuous) sum of harmonic oscillator hamiltonians (one for each normal mode), with an imaginary coupling.
With this in mind, in analogy with the harmonic oscillator studied in the previous section, can identify a metric operator
\begin{eqnarray}\label{eq:metric_QFT}
    \eta  =\exp\left\{-2\theta\int\,\frac{d^3p}{(2\pi)^3}\,
    \left[
    \tilde\varphi(-\vec p)\tilde\pi_a(\vec p) - \tilde a(-\vec p)\tilde \pi_\varphi(\vec p)
    \right]
    \right\},
\end{eqnarray}
where 
\begin{eqnarray}
    \tanh(2\theta) = \frac{2g}{m_A^2-m_\varphi^2},
\end{eqnarray}
which is precisely the value of the ansatz parameter $\theta$ that makes the hamiltonian (\ref{eq:field_hamiltonian_fourier_coupled}) $\eta-$hermitian, i.e.,
\begin{eqnarray}
 \hat\eta\hat H\hat\eta^{-1} = \hat H^\dagger.  
\end{eqnarray}

Just like in the quantum mechanical model (\ref{eq:hamiltonian-coupled-osc}), in the field theory (\ref{eq:model_lagrangian}) we can also define two important regions in the parameter space: (i) a {\it weak coupling regime}, for which $2g<|m_A^2-m_\varphi^2|$, so that $\theta\in\mathbb{R}$ and the spectrum of the hamiltonian is real and (ii) a {\it strong coupling regime}, for which $2g>|m_A^2-m_\varphi^2|$, where $\theta\in\mathbb{C}$ and the spectrum of the hamiltonian is complex. The latter is the ${\cal PT}$-broken phase and the critical value $g_{crit}=|m_A^2-m_\varphi^2|/2$ is an exceptional point. Note that the distinction between phases can be made at the level of the metric operator: a positive metric corresponds to the weak coupling regime, while a (generally) complex metric corresponds to the ${\cal PT}$-broken phase. Given the many subtleties associated with a nonhermitian metric, we will postpone the analysis of the ${\cal PT}$-broken phase to another work \cite{work-in-progress}.

In close analogy with the quantum mechanical case, we define the Schrödinger picture field operators
\begin{align}\label{eq:observable_fields}
    &\Phi(\vec x):= \eta^{-1/2}\varphi(\vec x)\eta^{1/2}\nonumber= \varphi(\vec x)\cosh\theta + ia(\vec x) \sinh\theta\nonumber\\
    &A(\vec x):= \eta^{-1/2} a(\vec x)\eta^{1/2}= -i\varphi(\vec x)\sinh\theta + a(\vec x)\cosh\theta,
\end{align}
and their respective conjugate momenta
\begin{align}\label{eq:observable_momenta}
    &\Pi_\Phi(\vec x):= \eta^{-1/2}\pi_\varphi(\vec x)\eta^{1/2}\nonumber= \pi_\varphi(\vec x)\cosh\theta + i \pi_a(\vec x) \sinh\theta\nonumber\\
    &\Pi_{A}(\vec x):= \eta^{-1/2}\pi_a(\vec x)\eta^{1/2}= -i\pi_\varphi(\vec x)\sinh\theta + \pi_a(\vec x)\cosh\theta,
\end{align}
which are $\eta-$hermitian operators, i.e., 
\begin{eqnarray}\label{eq:fields_eta_hermitian}
   \Phi(\vec x)^\#&=&\Phi(\vec x)\nonumber\\
    A(\vec x)^\#&=& A(\vec x)\nonumber\\ \Pi_\Phi(\vec x)^\#&=&\Pi_\Phi(\vec x)\nonumber\\
    \Pi_A(\vec x)^\#&=&\Pi_A(\vec x),
\end{eqnarray}
where the $\eta-$inner product and the  $\eta-$conjugation are defined exactly like in (\ref{eq:eta-inner-product}) and (\ref{eq:eta-conjugation-def}), respectively. Note that, due to the linearity of the Fourier transform and of the metric operator (which is here a constant operator throughout spacetime), perfectly analogous relations are valid for the Fourier-transformed (momentum-space) fields. Furthermore, note that the transformations (\ref{eq:observable_fields}) and (\ref{eq:observable_momenta}) both take the form of a rotation in the  $\varphi-a$ plane by an imaginary angle $i\theta$. 

In terms of these operators, the quantized field hamiltonian (\ref{eq:field_hamiltonian_fourier_coupled}) is given by
\onecolumngrid
\begin{eqnarray}\label{eq:field_hamiltonian_diagonal}
    H= \int \frac{d^3p}{(2\pi)^3} \left[
        \Pi_\Phi(-\vec p)\Pi_\Phi(\vec p) + \Omega_\Phi^2(\vec p)\Phi(-\vec p)\Phi(\vec p) + \Pi_A(-\vec p)\Pi_A(\vec p) + \Omega_A^2(\vec p)A(-\vec p)A(\vec p)
    \right],
\end{eqnarray}
\twocolumngrid
where 
\begin{align}
    \Omega_\Phi^2(\vec p) &= \frac{\cosh^2\theta\,\omega_\varphi^2(\vec p) + \sinh^2\theta\,\omega_a^2(\vec p)}{\cosh2\theta}=\vec p^2+M_\Phi^2,\nonumber\\
    \Omega_A^2(\vec p) &= \frac{\cosh^2\theta\,\omega_a^2(\vec p) + \sinh^2\theta\,\omega_\varphi^2(\vec p)}{\cosh2\theta}=\vec p^2+M_A^2,
\end{align}
so that we identify the masses of the observable fields (\ref{eq:observable_fields}) as
\begin{eqnarray}
    M_\Phi^2&=& \frac{\cosh^2\theta\,m_\varphi^2 + \sinh^2\theta\,m_a^2}{\cosh2\theta},\nonumber\\
    M_A^2&=& \frac{\cosh^2\theta\,m_a^2 + \sinh^2\theta\,m_\varphi^2}{\cosh2\theta}.
\end{eqnarray}

Note that the form (\ref{eq:field_hamiltonian_diagonal}) of the hamiltonian is manifestly self-adjoint with respect to the $\eta-$inner product, i.e., $\hat{ H}^\#=\hat{H}$. Besides, transformations (\ref{eq:observable_fields}) and (\ref{eq:observable_momenta}) also decomposes the the hamiltonian into two independent parts. In terms of the operators in (\ref{eq:observable_fields}) and (\ref{eq:observable_momenta}), the linear momentum operator (\ref{eq:linear_momentum}) is given by
\begin{eqnarray}
    {\vec {\cal P}} = \int d^3x\,\left[
    \Pi_\Phi(\vec x)\vec\nabla\Phi(\vec x) + \Pi_A(\vec x)\vec\nabla A(\vec x)
    \right]
\end{eqnarray}
which is also $\eta-$hermitian, i.e., ${\vec {\cal P}}^\#={\vec {\cal P}}$. 

Property (\ref{eq:fields_eta_hermitian}) ensures that both the hamiltonian and the linear momentum operators are observables of the field theory defined by the action (\ref{eq:model_lagrangian}). 

Given the form (\ref{eq:field_hamiltonian_diagonal}) of the hamiltonian,
it should be clear that it simply corresponds to a pair of decoupled scalar fields, which ultimately are equivalent to set of decoupled harmonic oscillators, one for each normal mode. Thus, one may expand the fields (\ref{eq:observable_fields}) and their conjugate momenta (\ref{eq:observable_momenta}) in terms of creation and annihilation operators, so that
\begin{align}\label{eq:observable_plane_wave_expansion}
    \Phi(\vec x)&=\int\frac{d^3p}{(2\pi)^3}\frac{1}{\sqrt{2\Omega_\Phi(\vec p)}}\left[
    \alpha_\Phi(\vec p)e^{i\vec p\cdot\vec x} + \alpha_\Phi^\#(\vec p)e^{-i\vec p\cdot\vec x}
    \right]\nonumber\\
    A(\vec x)&=\int\frac{d^3p}{(2\pi)^3}\frac{1}{\sqrt{2\Omega_A(\vec p)}}\left[
    \alpha_A(\vec p)e^{i\vec p\cdot\vec x} + \alpha_A^\#(\vec p)e^{-i\vec p\cdot\vec x}
    \right]\nonumber\\
    \Pi_\Phi(\vec x)&= -i\int\frac{d^3p}{(2\pi)^3}\sqrt{\frac{\Omega_\Phi(\vec p)}{2}}\left[
    \alpha_\Phi(\vec p)e^{i\vec p\cdot\vec x} - \alpha_\Phi^\#(\vec p)e^{-i\vec p\cdot\vec x}
    \right]\nonumber\\
    \Pi_A(\vec x)&= -i\int\frac{d^3p}{(2\pi)^3}\sqrt{\frac{\Omega_A(\vec p)}{2}}\left[
    \alpha_A(\vec p)e^{i\vec p\cdot\vec x} - \alpha_A^\#(\vec p)e^{-i\vec p\cdot\vec x}
    \right],
\end{align}
where
\begin{eqnarray}
    \left[
    \alpha_\Phi(\vec p),\,\alpha^\#_\Phi(\vec k) 
    \right] = (2\pi)^3\delta(\vec p+\vec k).
\end{eqnarray}
Note that the creation operators are given by $\alpha^\#(\vec p)$, rather than $\alpha^\dagger(\vec p)$. The hamiltonian (\ref{eq:field_hamiltonian}) can then be cast in the usual form
\begin{align}
    {H} &= \int\frac{d^3p}{(2\pi)^3}\left[
     \Omega_\Phi(\vec p)\left(\alpha_\Phi^\#(\vec p)\alpha_\Phi(\vec p)+\frac{\delta(\vec p=\vec 0)}2\right) \right.
     + \nonumber \\
     &\left. +\Omega_A(\vec p)\left(\alpha_A^\#(\vec p)\alpha_A  (\vec p)+\frac{\delta(\vec p=\vec 0)}2\right)
    \right],
\end{align}
so that the spectrum of the theory is the same as the one of a free theory of two real scalar fields. In terms of the creation and annihilation operators, the linear momentum operator is given by
\begin{eqnarray}
    {\vec{\cal P}}&=&\int\frac{d^3p}{(2\pi)^3}\vec p\left[
    a_\Phi^\#(\vec p)a_\Phi(\vec p)+a_A^\#(\vec p)a_A(\vec p)
    \right].
\end{eqnarray}

With all these properties being almost trivially identical to the ones of a free theory, the existence of a Fock space for the nonhermitian theory (\ref{eq:model_lagrangian}) is also an immediate consequence. The vacuum state is the one annihilated by the operators $a_\Phi$ and $a_A$, and any excited state is obtained from the vacuum by sucessive applications of the creation operators $a_\Phi^\#$ and $a_A^\#$. For example, denoting the vacuum state by $\ket{0}$, the state $a_\Phi^\#(\vec k)\ket{0}$ is a state of a single $\Phi$ particle with momentum $\vec k$ and energy $E_1=\sqrt{\vec k^2 + M_\Phi^2}$. All other excited states (i.e., the Fock space) can be built in a similar fashion.

One could now argue that the whole argument made so far is quite trivial, since the theory with imaginary coupling ultimately boils down to a theory of decoupled free fields with real parameters. To some extent, we cannot fully disagree with this remark, as is evident from the simplicity of the discussion above. However, this apparent triviality may be not so evident when one deals exclusively with correlation functions of some quantum field theory in a nonperturbative regime. This is the case, e.g., of functional methods such as the Schwinger-Dyson equations \cite{Alkofer:2000wg,Aguilar_2020,Fischer:2020xnb,Kondo_2020} or the Functional Renormalization Group \cite{Cyrol_2016,Cyrol:2018_SciPost} in the context of Yang-Mills theories. In such approaches, propagators with associated not-everywhere positive spectral functions are found.

In the next subsection, we will resume our discussion of Secs. \ref{sec:spectral-functions} and \ref{sec:coupled-oscillators} about two-point correlation functions, but now for the case of fields. As we will see, positivity violation of the K\"allen-Lehmann spectral function can be seen as a consequence of choosing operators that do not correspond to observables of the theory.% (even in a strictly perturbative sense, which is evidently the case of the free theory considered here). 

%%%%%%%%%%
\subsection{Two-point correlation functions and positivity violation}

Let us now turn our attention to time-ordered expectation values of products of field operators, i.e., two-point correlation functions, or simply propagators. We define the propagator
\begin{eqnarray}\label{eq:def_propagator_eta}
    D^{\Psi_1\Psi_2}_\eta(x,y):={\langle \psi_0\vert \,T\bigl \{ \hat \Psi_1(x) \hat \Psi_2(y)\bigr\}\,\psi_0\rangle}_\eta 
\end{eqnarray}
where $\ket{\psi_0}$ is the vacuum state, $T$ is the usual time-ordering operator, $\Psi_1$ and $\Psi_2$ are generic Heisenberg picture field operators, $x$ and $y$ are arbitrary spacetime points, and $\eta$ is the metric operator. The Heisenberg picture operators $\hat \Psi_H(x)$ can, as usual, be obtained from their Schr\"odinger picture counterparts $\hat \Psi_S(\vec x)$ as
\begin{eqnarray}\label{eq:heisenberg_picture_fields}
 \hat \Psi_H(x) = e^{i\hat Ht}\hat\Psi_S(\vec x)e^{-i\hat Ht},
\end{eqnarray}
so that $\hat \Psi_S(\vec x)=\hat\Psi_H(x)$, with $x=(t=0,\vec x)$.
In what follows, we shall omit the subscripts $H$ or $S$, unless the operator picture is not clear from the context. 

It is interesting to note that, since the hamiltonian is $\eta-$hermitian, $\hat H^\#=\hat H$, a given Schr\"odinger picture $\eta-$hermitian operator $\Psi_S(\vec x)=\Psi_S^\#(\vec x)$ will give rise to a corresponding $\eta-$hermitian {\it Heisenberg picture} operator. In other words, self-adjointness is preserved under time evolution. This is what happens with the field operators $\hat \Phi=\hat\Phi^\#$ and $\hat A=\hat A^\#$. On the other hand, any operator which is not $\eta-$hermitian, even if it is self-adjoint with respect to some other inner-product, will loose this property under time evolution. This is precesely what happens with the field operators $\hat\varphi=\hat\varphi^\dagger$ and $\hat a=\hat a^\dagger$.

Note that the definition of the propagator (\ref{eq:def_propagator_eta}) depends explicitly on the choice of the metric operator. In a hermitian field theory, one has, of course, $\hat \eta=\mathds{\hat 1}$, so that the metric does not appear in the calculation of the correlation functions. Note also that, in a path-integral formulation, the two-point function calculated in the usual way (i.e., by taking functional derivatives of the generating functional with respect to sources) will lead precisely to (\ref{eq:def_propagator_eta}), as discussed, e.g., in \cite{Mostafazadeh_2010_07,Mostafazadeh_2007_76}.

As a consequence of the discussion of Sec. (\ref{sec:spectral-functions}), we can explicitly see that the correlation functions including the operators $\varphi$ and $a$ of the action (\ref{eq:model_lagrangian}) do not have an everywhere positive spectral function. This is the case because, even though these operators are hermitian at some given time (say, $t=0$), their time evolution (given by a nonhermitian hamiltonian) is such that their hermiticity is lost in general for $t\not=0$. On the other hand, the operators $\Phi$ and $A$, defined in (\ref{eq:observable_fields}) will be $\eta-$hermitian for every time $t$. Since the hamiltonian is also $\eta-$hermitian, propagators of such fields will have a positive spectral function. In the following, we calculate such correlation functions, in order to show our claims explicitly.

First, let us calculate the propagator of the $\eta-$hermitian fields $\Phi$ and $A$. We find 
\begin{eqnarray}\label{eq:prop_PhiPhi}
    D^{\Phi\Phi}_\eta(x,y)&=&{\langle \psi_0\vert \,T\bigl \{ \hat \Phi(x) \hat \Phi(y)\bigr\}\,\psi_0\rangle}_\eta \nonumber\\
    &=& \int\frac{d^4p}{(2\pi)^4}\frac{i}{p^2-M_\Phi^2+i\epsilon}e^{ip\cdot (x-y)}
\end{eqnarray}
and
\begin{eqnarray}\label{eq:prop_AA}
    D^{AA}_\eta(x,y)&=&{\langle \psi_0\vert \,T\bigl \{ \hat A(x) \hat A(y)
    \bigr\}\,\psi_0\rangle}_\eta \nonumber\\
    &=& \int\frac{d^4p}{(2\pi)^4}\frac{i}{p^2-M_A^2+i\epsilon}e^{ip\cdot (x-y)},
\end{eqnarray}
where, as usual, the limit $\epsilon\rightarrow0^+$ is implicit. Furthermore, the mixed propagator $D^{A\Phi}_\eta(x,y)$ vanishes identically. 
The Fourier transform of the above propagators is immediate from the expressions (\ref{eq:prop_PhiPhi}) and (\ref{eq:prop_AA}), and their spectral functions are given, respectively, by
\begin{eqnarray}\label{eq:spectr_PhiPhi}
    \rho^{\Phi\Phi}(s)=2\pi\delta(s-M_\Phi^2)
\end{eqnarray}
and
\begin{eqnarray}\label{eq:spectr_AA}
    \rho^{AA}(s)=2\pi\delta(s-M_A^2)
\end{eqnarray}
Note that both spectral functions (\ref{eq:spectr_PhiPhi}) and (\ref{eq:spectr_AA}) are non-negative, as expected from the general discussion in Sec. (\ref{sec:spectral-functions}).

Now, let us consider the propagators and spectral functions associated with the operators $\varphi$ and $a$. Using the results (\ref{eq:prop_PhiPhi}) and (\ref{eq:prop_AA}) we find
\onecolumngrid
\begin{eqnarray}\label{eq:prop_varphivarphi}
    D^{\varphi\varphi}_\eta(x,y)&=&{\langle \psi_0\vert \,T\bigl \{ \hat \varphi(x) \hat \varphi(y)\bigr\}\,\psi_0\rangle}_\eta \nonumber\\
    &=& \cosh^2\theta{\langle \psi_0\vert \,T\bigl \{ \hat \Phi(x) \hat \Phi(y)\bigr\}\,\psi_0\rangle}_\eta -\sinh^2\theta {\langle \psi_0\vert \,T\bigl \{ \hat A(x) \hat A(y)\bigr\}\,\psi_0\rangle}_\eta \nonumber\\
    &=& \int\frac{d^4p}{(2\pi)^4}
    \left[\cosh^2\theta
    \frac{i}{p^2-M_\Phi^2+i\epsilon}
    -
    \sinh^2\theta
    \frac{i}{p^2-M_A^2+i\epsilon}
    \right]e^{ip\cdot (x-y)},
\end{eqnarray}
and
\begin{eqnarray}\label{eq:prop_aa_}
    D^{aa}_\eta(x,y)&=&{\langle \psi_0\vert \,T\bigl \{ \hat a(x) \hat a(y)
    \bigr\}\,\psi_0\rangle}_\eta \nonumber\\
        &=& \cosh^2\theta {\langle \psi_0\vert \,T\bigl \{ \hat A(x) \hat A(y)\bigr\}\,\psi_0\rangle}_\eta-\sinh^2\theta {\langle \psi_0\vert \,T\bigl \{ \hat \Phi(x) \hat \Phi(y)\bigr\}\,\psi_0\rangle}_\eta \nonumber\\
    &=& \int\frac{d^4p}{(2\pi)^4}
    \left[\cosh^2\theta
    \frac{i}{p^2-M_A^2+i\epsilon}
    -
    \sinh^2\theta
    \frac{i}{p^2-M_\Phi^2+i\epsilon}
    \right]e^{ip\cdot (x-y)},
\end{eqnarray}
and also the mixed propagator
\begin{eqnarray}\label{eq:prop_avarphi}
    D^{a\varphi}_\eta(x,y)&=&{\langle \psi_0\vert \,T\bigl \{ \hat a(x) \hat \varphi(y)
    \bigr\}\,\psi_0\rangle}_\eta \nonumber\\
    &=& i\sinh\theta\cosh\theta \left[
     {\langle \psi_0\vert \,T\bigl \{ \hat \Phi(x) \hat \Phi(y)\bigr\}\,\psi_0\rangle}_\eta-
    {\langle \psi_0\vert \,T\bigl \{ \hat A(x) \hat A(y)\bigr\}\,\psi_0\rangle}_\eta \right]\nonumber\\
    &=& i\cosh\theta\sinh\theta\int\frac{d^4p}{(2\pi)^4}
    \left[
    \frac{i}{p^2-M_\Phi^2+i\epsilon}
    -
    \frac{i}{p^2-M_A^2+i\epsilon}
    \right]e^{ip\cdot (x-y)}.
\end{eqnarray}
\twocolumngrid
The corresponding spectral functions are given by
\begin{eqnarray}\label{eq:spectr_hermitian}
    \rho^{\varphi\varphi}(s)&=& 2\pi\left[
    \delta(s-M_\Phi^2)\cosh^2\theta-\delta(s-M_A^2)\sinh^2\theta
    \right],\nonumber\\
    \rho^{aa}(s)&=&2\pi\left[
    \delta(s-M_A^2)\cosh^2\theta-\delta(s-M_\Phi^2)\sinh^2\theta\right],\nonumber\\
    \rho^{a\varphi}(s)&=&2\pi i\sinh\theta\cosh\theta\left[
    \delta(s-M_\Phi^2)-\delta(s-M_A^2)\right].
\end{eqnarray}
It is clear that the spectral functions (\ref{eq:spectr_hermitian}) are not everywhere positive. This could be anticipated from the fact that the field operators $\varphi(x)$ and $a(x)$ are not self-adjoint, even though they are hermitian at $t=0$.

%%%%%%%%%%%%%%%%%%%%%%%%%%%%%%%%%%%%%%%%%%%%%%%%%%%%%%%%%%%%%%
\section{Discussion and a conjecture} 
\label{sec:discussion}

After discussing in detail a full solution for a quantum mechanical system of two harmonic oscillators coupled through an imaginary bilinear term, we analyzed the analogous case of two coupled scalar fields. The mathematical simplicity of the model allowed for explicit solutions, which exemplify generic features of pseudo-hermitian theories in a very transparent manner. 

We showed that, in a weak coupling regime, the spectra of both theories are real, in spite of the non-hermiticity of the hamiltonians. In particular, the hamiltonians are found to be \(\eta\)-pseudo-hermitian (with analogous metrics), and observable operators of the theory are also constructed from $\eta-$hermitian operators.

We believe that our study of two-point correlation functions may contribute to the discussion on the relation between positivity violation of the spectral functions, the non-observability of the respective operators, and the unitarity of the theory. With explicit examples, we could conclude that the non-self-adjointness of a given operator under time evolution leads to a non-positive (in general, complex) spectral function (given an inner-product related to the hamiltonian). 

%{\color{blue}For example, consider a (Schr\"odinger picture) hermitian operator $\hat A=\hat A^\dagger$, so that $\hat A(t)$ is its corresponding Heisenberg picture operator, so that $\hat A(0)=\hat A$. Now let $C_{\hat{\mathds{1}}}^{AA}$ be a time-ordered vacuum correlation function, as in (\ref{eq:generic_corr_funct}). If the time-dependent operator $\hat A(t)$ is hermitian at all times, then the corresponding spectral function will be positive. However, if the spectral function can be negative, this implies that, although $\hat A = \hat A(0)$ is hermitian, this is not true for $\hat A(t)$ in general. This, in turn, implies that the hamiltonian, being the generator of time translations, is itself not hermitian with respect to this inner product. Finally, a non hermitian hamiltonian implies that the time evolution operator is not unitary. }

In other words, if some spectral function is calculated in any quantum theory and it is found to be not positive-defined, we conjecture that this can be evidence for a nontrivial metric in Hilbert space. If such nontrivial metric is confirmed, it is also possible that an exceptional point exists somewhere in the theory's parameter space. Then, beyond such an exceptional point, a regime with complex energy eigenvalues (i.e., a ${\cal{PT}}-$broken phase) should appear. In sum, although we do not have a proof of this statement, we believe that positivity violation of the spectral function (even in the phase of real eigenenergies) can be a sign for the existence of a phase with complex energy eigenvalues. 

In particular, we conjecture that positivity violation in Yang-Mills theories, which is a well-known feature of the gluon propagator $\vev{A_\mu^a(x)A_\nu^b(y)}$ in the non-perturbative regime, can be a consequence of a nontrivial metric in Hilbert space. This idea will be put under scrutiny in an upcoming work \cite{work-in-progress-2}. 

In condensed matter physics we hope that our work will help to stablish reasonable experimental signatures for $\mathcal{PT}$-symmetry breaking for strongly correlated electronic systems. An interesting direction for future investigations is how the metric tensor constrains the spectral function of metallic systems with quadrupolar electron-electron interactions \cite{Aquino2020,Aquino2024} and if this would connect with measurements of the dynamical nematic response using methods such as time resolved resonant inelastic x-ray scattering and momentum-resolved electron energy-loss spectroscopy. 

The field theoretical model discussed in this paper corresponds to a free theory. However, similar methods can be applied to interacting theories. For example, the lagrangian density
\begin{align}\label{eq:O2-interaction}
    {\cal L}_{O(2)} &= \frac12\partial_\mu\varphi\partial^\mu\varphi - \frac{m_\varphi^2}{2}\varphi^2 + \frac12\partial_\mu a\partial^\mu a - \frac{m_a^2}{2}a^2 \nonumber \\
    &- ig\varphi a
    - \lambda\left(\varphi^2+a^2\right)^2,
\end{align}
is pseudo-hermitian and its metric is exactly given by (\ref{eq:metric_QFT}). Since it is in general very difficult to find a closed expression for the metric operator (especially in QFT), the theory (\ref{eq:O2-interaction}) can be an interesting toy model for future studies of interacting theories. Of course, infinitely many other interacting pseudo-hermitian theories are possible, with different metric operators.

Evidently, an important regime of the theory that was not explored in this work is the {\it strong coupling} regime (defined in Sec. \ref{sec:coupled-oscillators}), where the energy eigenvalues are complex and the metric operator is not positive. Actually, we believe that this regime is the most interesting phase of the theory, with many subtleties involved. We will postpone this analysis (both in quantum mechanics and in quantum field theory) to a future work.

%%%%%%%%%%%%%%%%%%%%%%%%%%%%%%%%%%%%%%%%%%%%%%%%%%%%%%%%%%%%%%
%%%%%%%%%%%%%%%%%%%%%%%%%%%%%%%%%%%%%%%%%%%%%%%%%%%%%%%%%%%%%%

\section*{Acknowledgements}

The work of B.W.M. is supported by the Brazilian agencies CNPq, CAPES, FAPERJ and FAPESP. I.Y.P. is supported by a CAPES MSc. grant. This study was financed in part by the Coordenação de Aperfeiçoamento de Pessoal de Nível Superior – Brasil (CAPES) – Finance Code 001. R.A. is supported by a Post-Doctoral Fellowship No. 2023/05765-7 granted by São Paulo Research Foundation (FAPESP), Brazil.

%%%%%%%%%%%%%%%%%%%%%%%%%%%%%%%%%%%%%%%%%%%%%%%%%%%%%%%%%%%%%%
%%%%%%%%%%%%%%%%%%%%%%%%%%%%%%%%%%%%%%%%%%%%%%%%%%%%%%%%%%%%%%

\newpage
\bibliography{PTSymm-Biblio}

%%%%%%%%%%%%%%%%%%%%%%
%%%%%%%%%%%%%%%%%%%%%%
%%%%%%%%%%%%%%%%%%%%%%

%\input{notes.tex}

%%%%%%%%%%%%%%%%%%%%%%
%%%%%%%%%%%%%%%%%%%%%%
%%%%%%%%%%%%%%%%%%%%%%

\end{document}